# Simple three-stage frequency-stabilized diode laser system using injection-locking and tapered amplifier

J. Szonert, M. Głódź, and K. Kowalski

*Institute of Physics, Polish Academy of Sciences,*
*Al. Lotników 32/46, 02-668 Warsaw, Poland*

**Abstract:** We developed a simple three-stage amplifying tunable cw diode laser system that comprises: a Littrow configuration master laser, frequency stabilized with a dichroic atomic vapor laser lock (DAVLL), an acousto-optic frequency shifter (AOM), an injection-locked slave laser and a tapered amplifier (TA). Slave amplifies the frequency-shifted beam and suppresses (within 0.5%) the strong dependence of its intensity on the AOM carrier frequency, thus acting as an efficient optical limiter. The beam is further amplified in TA. System operates at 780 nm, with an output power above 700 mW, time-averaged linewidth of 0.6 MHz, and a frequency drift below 2 MHz/h. The mode-hop-free tuning range of the master amounts 2 GHz, DAVLL allows a lock point tunability in the range of 400 MHz. The fine (with some tens of kHz precision) tuning range spans 70 MHz, determined by the AOM model applied. A detailed description of the system was given and its performance was tested. The basic components were designed and manufactured in our lab.

## 1  Introduction

Tunable continuous-wave (cw) frequency stabilized lasers of narrow linewidth and sufficient optical power are required in diverse applications, e.g., precision spectroscopy or laser cooling and trapping. Diode lasers (DLs) are compact, cost-effective devices available at wide range of emission wavelengths, from mid-infrared to violet, that since their presentation in 1962 have undergone enormous technological progress and have become ubiquitous source of laser radiation in many areas [1]. However, conventional solitary single-mode free-running narrow-waveguide Fabry-Perot (FP) DLs fail to meet the above requirements. Short cavities of low finesse, broad gain bandwidths and inherent instabilities cause broadband emission (tens of MHz); frequency stability and tunability achieved solely by altering the temperature and/or current is coarse and suffers from mode-hops; while the power density related facet degradation limits the output power [2,3]. Various approaches are used to improve performance of DL-based optical sources, some essential ones related predominantly to FP DLs are briefly sketched below.

The extreme sensitivity of DL to optical feedback is used in the extended (external) cavity diode laser (ECDL), with cavity length and feedback strength being the primary factors influencing its performance [4,5]. Optical components, such as various types of gratings (reflective [6], transmission [7,8] or volume [9-11]), etalons [12], filters of a narrow [13] or wide [14] bandwidth, plane mirrors [15] and glass-slips [16], are utilized to produce frequency-selective optical feedback, which narrows linewidths, enhances stability and enables tunability. Alternatively, feedback from the high-finesse optical cavity [17], a combination of ECDL and optical cavity [18] or a compact whispering-gallery-mode resonator [19,20] are used, which reduces the linewidth to the level of kHz or even Hz. By far the most common are ECDLs based on reflective gratings in Littrow [6,21,22] or Littman-Metcalf [23,24] configurations, each with its own advantages.

Issues related to ECDL have been extensively investigated. To extend the mode-hop-free (MHF) tuning range various approaches are employed such as: optimization of pivot position [24,25]; adjustment of the cavity length or grating offset to match the rates of grating rotation and cavity tuning [26]; synchronization of the grating position with the bias current (feedforward) [16,27,28]; the use of multiple actuators for the matched translation and rotation of the grating [7,28,29]; active control of the cavity length by means of a feedback loop based on polarization spectroscopy [30] and power monitoring [31]; or the common use of an antireflection (AR) coating on the DL output facet [32]. However, MHF range of 137 GHz was obtained for a short cavity Littrow type laser supplied with a standard non-AR-coated DL, when the current and the grating position changed synchronously [33]. Other studies include, e.g., optimization of power and spectral properties by means of a wave-plate placed inside the cavity [15,34]; impact of the collimator position on the feedback efficiency and linewidth [35]; or dependence of the linewidth on the grating's properties [36,37]. Factors both fundamental and ambient induced determine the effective linewidth and stability in ECDLs. Thermal, mechanical, acoustic, bias current and air pressure disturbances can be a source of linewidth broadening, intensity fluctuations and frequency drift [35,38,39]. The countermeasures can be divided into passive and active.

For the former a careful, compact, stable mechanical design and decoupling from environmental perturbations is used [7,21,29,38,40,41]. For example, an ECDL design with superior passive characteristics is presented in [42], while the novel concept of a compact micro-integrated tunable ECDL, with no moving parts, suitable for precision spectroscopy, is demonstrated in [10]. High sensitivity of the optical frequency to bias current and temperature (~ -3 MHz/μA and ~ -8 MHz/mK, respectively [35,38,39]) requires the use of precise controllers. While those on the market often meet high expectations, many groups develop their own current [2,43-48] and temperature [2,49,50] controllers in response to special needs or for budgetary reasons.

Numerous techniques have been developed to stabilize both short and long term laser frequency fluctuations. They differ in underlying physical mechanisms, performance, sensitivity to external perturbations and complexity [51-54]. In the present work a dither-free dichroic atomic vapor laser lock (DAVLL) technique is applied [51,55,56], for its broadly tunable locking range, simplicity and reliability. (Note: Variants thereof are in use, including the Doppler-free approach [57-59] or the related Faraday rotation scheme [60].)

Typically a complete ECDL based arrangement for high resolution applications provides a spectrally and spatially pure beam with a linewidth of tens to several hundred kHz and MHF tunability of up to tens of GHz, but its output power rarely exceeds tens of mW, partly due to losses in many optical components. However, technological advances have made it possible to generate many Watts from a single emitter [1]. For broad-area lasers (BALs), the width of the gain region ranges from tens to hundreds of μm, which compares with just a few μm in a standard narrow-waveguide device. But the high power comes at the expense of broad (up to several nm) spectral bandwidth, and poor beam quality. Tapered diode lasers (TLs), and closely related tapered amplifiers (TAs), combine the advantages of low power (narrow-waveguide) and high power (broad-waveguide) diode lasers, resulting in better stability, improved spectral characteristics and a nearly diffraction limited beam quality [61]. In more refined systems various extended cavity configurations combine with BAL [62-64], TL [65,66] or TA [13,24,67], providing in addition to high power also tunability, improved stability and reduced linewidth. In some implementations, spectrally narrow and nearly Gaussian beam is obtained [13,24].

An efficient approach to producing a high power and high spectral quality laser beam is to separate the generation of the spectrally narrow and diffraction limited beam from its amplification. There are three basic schemes that use semiconductor gain devices [68]: (i) resonant amplification by the optical injection-locking (OIL) [69,70], (ii) double-pass amplification in the BAL (with an AR-coated front facet and an input beam injected at a small angle (V-injection) [71,72] and (iii) travelling-wave amplification in TA (with both facets quality AR-coated) [68,73-77]. Two latter schemes are known as the master oscillator power amplifier (MOPA).

In this paper we present a DL-based cw optical source where both OIL and MOPA approaches, often used separately, are applied simultaneously in a three-stage, amplifying laser system using TA as the power unit. The system designed and built in our lab has an output power above 700 mW and a wavelength of 780 nm. The paper is organized as follows. In the introductory part, optical sources based on Fabry-Perot diode lasers are briefly presented and various approaches to improve their performance are given. Then, a detailed description of the structure and properties of each subsystem of our laser setup is given. Interesting features are detailed where appropriate. Some performance tests and measurements are included. Finally a brief conclusions are given.

## 2 Overview of the system

For experiments with cold Rb atoms in a MOT an intense cw single-frequency, stabilized and tunable laser source of narrow-linewidth and wavelength of 780 nm, at the Rb D2 transition, is required. We have developed a DL-based three-stage amplifying system which consists of (a) a low-power narrow linewidth ECDL (master laser), followed by two amplification stages applied in series: (b) an OIL amplifier (slave laser) and (c) a TA power unit. The master laser is frequency-locked with the DAVLL system. Before entering the OIL (slave) amplifier, the master beam is frequency-shifted by an acousto-optical modulator (AOM), which allows the output beam to be tuned. The arrangement is completed by diagnostics tools. The optical layout of all functional subsystems is shown in Fig. 1. In what follows they are described in detail.

### 2.1 Master laser

Our grating-stabilized ECDL in Littrow configuration (master, DL1 in Fig. 1) follows, with some modifications, the design of Hänsch and co-workers [21]. It includes a popular (non-AR-coated), edge-emitting single-longitudinal mode DL (Sanyo DL7140-201M), an aspheric collimator (Thorlabs C230TME-B, f = 4.5 mm, NA = 0.55, AR-coated) and a diffraction grating (Thorlabs GR13-1850, 1800 mm$^{-1}$). Although the laser used is of type

'off-the-shelf', it was selected for the appropriate wavelength from a batch consisting of several pieces. Main components of the laser assembly are machined of brass. Other materials like, e.g., 'neusilber' alloy [21] may provide better mechanical or thermal properties, but were unavailable at the time of construction. The diode and the lens are fixed in a single mount in a way that secures good thermal contact and allows for fine adjustment of collimation. The major axis of the elliptical beam is set horizontal and the laser polarization is vertical. The grating with groves set vertical is glued to the arm of a flexible hinge (set at ~ 45º to the beam) and can be tilted horizontally with a fine-pitch screw. For a fine frequency tuning a low voltage piezoelectric transducer (PZT, Piezomechanik Pst 150) is placed between the rear surface of the arm with the grating and the tip of the screw (secured with a ceramic disk).

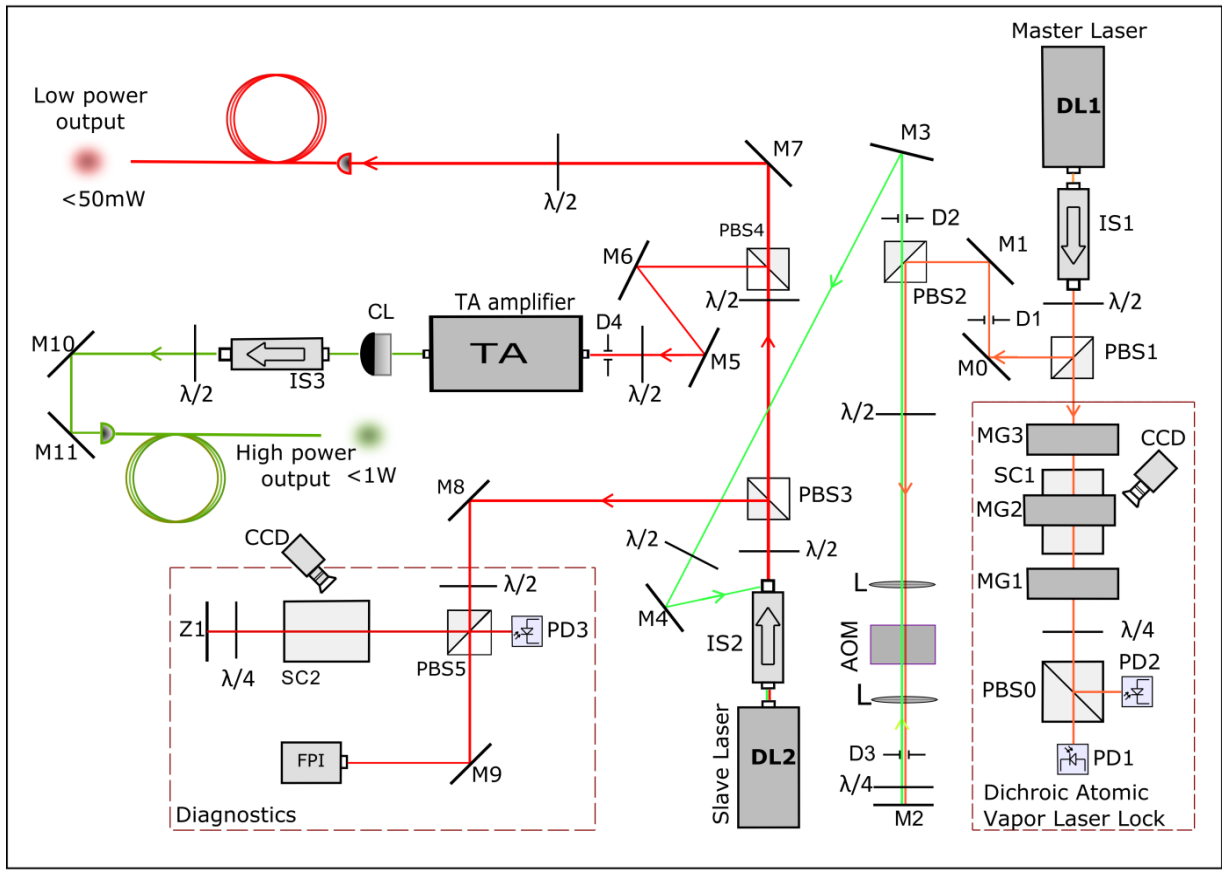

**Fig. 1** Optical layout of the amplifying laser system. DL1 – master diode laser, DL2 – slave diode laser, TA – tapered amplifier, PBS(0-5) – polarizing beam splitter, M(1-11) – dielectric mirror, Z1 – partially reflecting mirror, λ/2 (λ/4) – half (quarter) -wave plate, PD(1-3) – photodiode, IS(1-3) – optical isolator, D(1-4) – aperture, MG(1-3) – permanent torroidal magnet, L – lens, CL – cylindrical lens, AOM – acousto-optical modulator, SC(1-2) – Rb vapor cell, FPI – scanning interferometer, CCD – camera.

The vertical axis of the grating can be tilted with another fine-pitch screw, by flexing the surface of the flexure mount (size 40 x 80 x 10 mm$^3$) to which the grating's mount is rigidly bolted. The first order diffracted beam (~ 15–20% efficiency) reflects back into the DL cavity, while the zeroth order beam forms the output. A relatively short cavity (~ 15 mm) reduces the mode competition and improves mechanical stability. A beam steering mirror (8 x 6 x 1 mm$^3$) fixes the output beam direction [78], however, a small parallel beam walk still persists. Installed near the DL mount is a PCB board with an anti-spike protection circuit and a relay that shorts a diode when not in use, and allows safe disconnection of cables. An on-board jumper selects the DL class: cathode- or anode- grounded (DL used belongs to the first category). PCB contains also a bias-tee component enabling high frequency modulation of the current (option not used in this application), it is also equipped with appropriate connectors: two D-sub 9-pin and two BNC ones, the first two for connection to the current and temperature controllers, and the second two for PZT control and current modulation, respectively. For temperature readout, the LM35 sensor is glued inside the DL mount with a thermally conducting adhesive. However, we found later that better short term stability of ±0.1 mK is achieved with the 10 kΩ NTC thermistor. A thermoelectric cooler (TEC, 40 x 40 mm$^2$) is sandwiched between the optical cavity mounting plate and the aluminum base (80 x 120 x 25 mm$^3$) and fixed with plastic screws. The ECDL is protected by an aluminum housing to reduce the ambient disturbances. The base is fixed to a massive heat sink, also made of aluminum, which is clamped to the optical breadboard. The temperature is stabilized typically at 17-20 °C with the lab-constructed precision PID controller (compatible with LM35, 10 kΩ NTC and AD590 sensors) with a 6 h stability of ±0.2 mK [79]. The laser is powered by a microprocessor-based high-precision controller designed and built in the laboratory [80]. It is characterized by high long-term stability (±50 nA in 6 h) and very low ripple and noise (< 50 nA rms in 10 Hz − 300 kHz range; both values are limited by the sensitivity of the available test equipment), and provides a knob adjustable current of 0 − 350 mA, switchable in four bands. The design incorporates several safety features: slow start/fast stop (the latter to protect the DL in case of power drop out), adjustable current and voltage limiters, filters, relay control. It also supplies the tunable voltage (±15 V range) used to the PZT steering. This voltage can be modulated externally. (Note: the frequency response of PZT actuators is limited to ~10 kHz, however, due to the combined mass and stiffness of the grating arm the effective bandwidth is reduced to hundreds of Hz. We typically use an asymmetric triangular wave at 10-30 Hz. It was found that its falling (return) slope must be longer than 2 ms to avoid mechanical resonance and the resulting distortions.) The design enables also (i) internal (DC to 6 kHz) or external (up to 2 GHz) current modulation with depth up to 10% of its full scale range and (ii) current correction synchronous with PZT voltage (feedforward compensation).

ECDL is typically biased well above threshold (~ 2–3 $I_{th}$) to narrow the linewidth and generate sufficient power. The averaged (within a few seconds) linewidth is estimated to be about 0.6 MHz, based on laser tuning to the steep slope of the atomic resonance, and relating fluctuations of the transmitted power to the optical frequency noise. Independently, in a preliminary optical heterodyne beat signal measurement of two similar lasers detected by a fast photodiode, the linewidth below 1 MHz was determined. No special efforts were made to obtain the long MHF tuning range, hence the modest value of ~ 0.5 GHz was obtained with the PZT tuning only. However, with simultaneous current corrections (with feedforward coefficient $\Delta I/\Delta V \approx$ - 0.5 mA/V), the range was extended to ~ 2 GHz.

The linearly polarized beam is sent through an optical isolator (Isowave I-80-5M-SD, 60 dB, transmission ~ 75%) to protect against back-reflection, and then split by a polarizing beam splitter (PBS1), which together with a half-wave plate acts as a variable beam divider.

The reflected beam of maximum power ~ 12–15 mW constitutes the output, whereas the small transmitted fraction is used for frequency stabilization.

## 2.2 DAVLL frequency stabilization system

Locking the single-frequency laser to a narrow resonance is a prerequisite for all high-precision experiments. In the DAVLL scheme a magnetic field splits the Doppler-broadened lines and the signal of the induced circular dichroism is utilized as an error signal [55,56,81,82]. The scheme of a circular polarimeter is shown in Fig. 1. A weak (~ 0.1 mW) linearly polarized beam passes through a home-made cylindrical glass cell (of 6 cm length, 3 cm diameter) with natural Rb vapor. Three identical torroidal permanent magnets, with an outer (inner) diameter 15(12) cm, thickness of 1.5 cm and 4 cm spacing, surround the cell and produce a reasonably uniform (~ 5%) axial magnetic field of 130 Gauss, which is close to optimal values for the Rb D2 transitions [51]. Circular polarization components $\sigma^{\pm}$ generate Zeeman absorption profiles displaced symmetrically in opposite directions in frequency. After conversion (with quarter-wave plate) and spatial separation (with PBS0) the orthogonal linearly polarized beam components are detected with photodiodes (PD1, PD2). Subtraction of both profiles by the low noise differential amplifier provides a broad dispersion signal (with a monotonic slope) which is compared with the reference voltage. The locking point can be tuned continuously, either optically by rotating the quarter-wave plate, or (more precisely) electronically by changing the reference voltage. The resulting signal is then fed to an external PID based servo, then the feedback voltage is applied to the PZT input of the laser controller. The detecting part of the setup is enclosed in the case to keep both photodiodes at the same temperature and to protect them from ambient light. Apart from the advantages such as the simplicity of a dither-free scheme, a wide capture range, and a broad continuous locking frequency range of ~ 400 MHz around the Rb D2 transitions, this method suffers also from some flaws. Vapor pressure-temperature dependence, quality of optical components, drifting electronics, and beam intensity noise contribute to the lock point instabilities, necessitating periodic adjustments. However, ability to stabilize the laser at an arbitrary detuning, robustness and simplicity of configuration are the greatest merits of this method. For coarse control, fluorescence is monitored with a CCD camera. The frequency calibration is independently performed using the sub-Doppler reference spectrum, as discussed below. Under typical circumstances, for lock point set in vicinity of $^{85}$Rb $5S_{1/2}$ (*F*=3) → $5P_{3/2}$ (*F'*) transitions, without room temperature stabilization and special efforts, a slow drift below 0.5 MHz/15 min is observed.

## 2.3 AOM frequency shifter

Acousto-optic modulators (AOMs) are used in applications where precise sweeping of the laser frequency or its on demand setting at a fixed detuning is required. Besides ability to control the frequency and intensity AOM deflects the beam by an angle dependent on its RF driving frequency. When frequency sweeps or jumps are foreseen the undesired angular variations create problems with alignment of the setup. A solution is offered by the AOM in a double-pass configuration which doubles the frequency shift, broadens the tuning bandwidth and to significant extent compensates the beam pointing fluctuations [83]. Note that higher multi-pass schemes, odd [84] or even [85], high diffraction orders [86] or combinations thereof [87], can be employed to expand the range of frequency shifts.

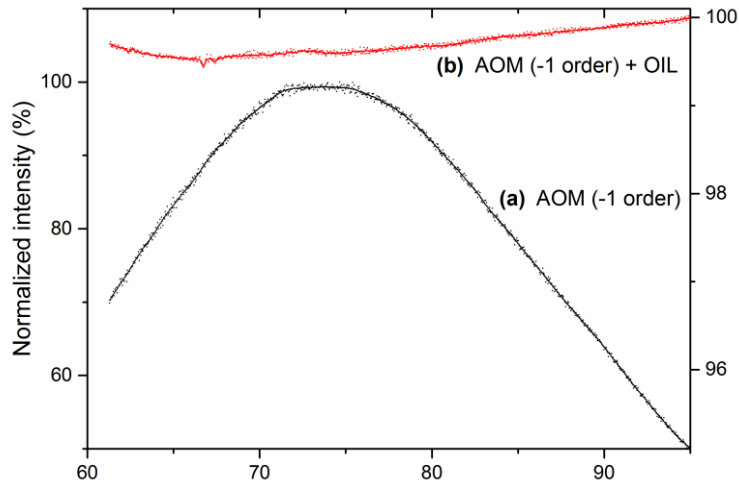

**Fig. 2** Relative response of the AOM (a), and the AOM followed by the OIL (b), as a function of RF carrier frequency. Double-pass AOM is aligned in the -1 order, inducing the optical frequency down-shift. (a) Response of the double-pass AOM only (lower trace, black curve, left vertical scale); (b) Combined response of the AOM followed by the OIL stage (upper trace, red curve, right vertical scale). The optical limiting of the OIL amplifier is revealed.

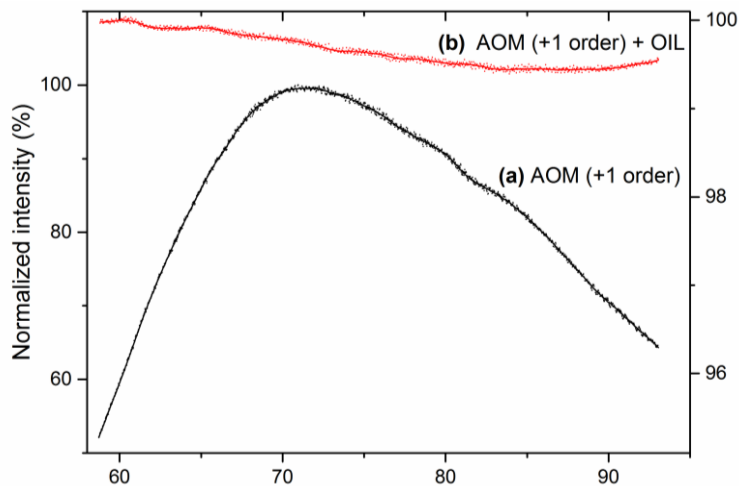

**Fig. 3** The same as in Fig. 2, but here the double-pass AOM is aligned in the +1 order, inducing the optical frequency up-shift. The optical limiting of the OIL amplifier is revealed.

We use the AOM (Crystal Technology 3080-125, 80 MHz center frequency, 25 MHz specified bandwidth) with a dedicated driver (1080AF) based on a VCO (voltage controlled oscillator). The driver is supervised by the laboratory-made digitally controlled source providing: (i) a tuning voltage (0 − 15 V; for convenience calibrated in MHz) that allows a precise tuning of the RF frequency in absolute or relative increments, and (ii) a variable control voltage (0 − 1 V) for amplitude adjustment. An optical setup includes two f = 160 mm lenses (non-AR-coated, thus inducing losses), two wave-plates and a mirror (M3). The half-wave plate matches the polarization of the beam with the axis of the AOM crystal. The quarter-wave plate transforms the polarization so that the twice diffracted beam can be separated from the input beam on PBS2. By proper alignment the -1 (+1) diffraction order is selected in each pass to accomplish optical frequency down-shift (up-shift, respectively). Spurious beam orders are filtered out with the apertures (D2, D3). The relative response of the double-pass AOM versus RF carrier frequency is depicted in Fig. 2(a) for -1 order and in Fig. 3(a) for +1 order. While dependencies look similar the optical frequencies are shifted by the doubled RF frequency to the red in Fig. 2(a) and to the blue in Fig. 3(a), respectively. Substantial intensity variation and asymmetry with respect to the specified center value is seen.

The peak diffraction efficiency for both configurations is ∼ 40%, while the RF precision related mainly to VCO stability is several tens of kHz. Thus, without compromising stability of the laser frequency lock, offset centered at +160 MHz (or -160 MHz) and tunable in the bandwidth ∼70 MHz, sums up with the master beam frequency. Although quite narrow, this bandwidth covers interesting features of the Rb spectrum. However, combined losses reduce the available peak power to ∼ 2 mW only.

## 2.4 Optical injection-locking

Optical injection of DLs has been investigated and utilized since 1980s [69, 70,88,89]. In a stable optical injection-locking (OIL) regime, the slave laser replicates the spectral and spatial properties of the injected weak seed beam from the master laser and amplifies its power. Two parameters are important for OIL: the injected power $P_{inj}$ and the detuning $\Delta v = v_m - v_s$ between the slave ($v_s$) and master ($v_m$) laser frequencies. Stable OIL occurs for small, most often negative, detunings and for $P_{inj}$ above the locking threshold. For $P_{inj}$ held constant, the span of detuning frequency at which the stable OIL occurs determines the locking range. The latter broadens and redshifts as $P_{inj}$ increases.

In our setup the slave laser (DL2 in Fig.1) and its collimator are held in a collimation tube (Thorlabs LT230P-B) fixed in a solid aluminum block and mounted via a TEC on the heat sink. The laser is current and temperature stabilized and enclosed in its aluminum housing. To initiate the OIL a few requirements have to be addressed:

(i) *Mode and polarization matching, alignment.* To successfully inject the seed beam into the slave aperture, both laser beams need to be of similar size and shape. This condition is relaxed by the fact that DL2 and its collimator are of the same type as their counterparts in the master module. While more accurate mode matching could result in stronger coupling, no additional beam shaping was performed. The single stage optical isolator (Isowave I-7090-CH, 40 dB , ∼ 85% transmission) both allows optical injection through its escape port, and protects the slave from back-reflections and backward emission of TA. The half-wave plate is used to maximize the isolator's throughput and to match the polarization direction of the seed with that of DL2. The overlapping of the seed and slave beams is done by adjusting two mirrors (M3, M4) and positioning the collimator. Due to the small size of the front facet, careful monitoring of the light entering the DL2 waveguide is required. We observed that the voltage build up in the laser junction operated in a photovoltaic (zero-bias) mode provides useful criterion of injection efficiency. Therefore, the controller has a built-in dedicated add-on feature that measures the voltage directly at the junction when the protective relay remains open and the bias current is off. With a seed power of up to ~ 2 mW (measured upstream of the isolator's escape port) the signal spans 0.7 – 2.2 V, providing a sensitive optimization method. In this approach, there is no need to open the case, disconnect or re-wire the DL2 before measurements. Alternatively, the voltage drop across the resistor (10 kΩ) connected in parallel with the junction was measured in the convenient 0 – 1.4 V range. However, the resistor has to be connected very carefully due to the risk of the DL2 failure. Once the OIL is provisionally established, the master power is reduced and the seed beam is re-aligned by maximizing injection efficiency. (While preparing the manuscript, we found that a similar approach, but with DL disconnected from the controller, was used in [90].)

(ii) *Injection level and frequency matching.* The free-running DL2 biased far above the threshold at ~ 80 – 120 mA (~ 2–3$I_{th}$) emits ~ 50 mW as measured behind the isolator. The injection power measured at the escape port has a range of ~ 0.8 – 2 mW for different AOM arrangements and RF frequencies, however, the actual injection levels may have a higher dynamic range due to the residual beam pointing variation. No efforts were undertaken to determine exact amount of the light coupled into the waveguide, but it has proven adequate for stable OIL in the applied working conditions. With master's working point kept fixed the slave frequency is tuned by adjustments of the temperature and the bias current until frequencies $\nu_m$ and $\nu_s$ match quite closely. Typically OIL is accomplished within narrow (< 0.5 mA) current windows separated by ~10 – 15 mA intervals, which facilitates optimization of both the output power and the locking range. (Similar quasi-periodic behavior was investigated in [91].) The latter is done by changing the slave current (and temperature) while sweeping the master frequency and monitoring the stability of the OIL with a diagnostic system.

(iii) *Diagnostics.* Part of the beam split off with a variable beam divider (a half-wave plate and PBS3) is sent to the saturated absorption spectroscopy (SAS) system in a simple single-probe-beam configuration. It consists of a room temperature Rb vapor cell (SC2, Thorlabs GC19075-RB, 75 mm length), partially reflecting mirror (Z1, ~10 %), quarter-wave plate, polarizing beam splitter (PBS5) and photodiode (PD3). The frequency of the beam is swept by PZT or AOM. The SAS signal allows frequency calibration using known sub-Doppler features as a reference [6,92,93], helps extend the locking range and provides hints on frequency stability and spectral purity. Unstable or multi-mode operation is revealed as a shivering or noisy spectrum. The particular role of SAS is to establish the origin for AOM frequency sweeps, by offset tuning the master frequency relative to the selected spectral feature. As an auxiliary tool, the CCD camera enables continuous monitoring of Rb fluorescence. A fraction of the beam is fed to a scanning FP spectrum analyzer (Toptica FPI 100, FSR=1 GHz, finesse > 500) to monitor the slave spectrum and evolution of the OIL. When the slave is off, the seed beam is partially reflected from the facet and can also be analyzed. Thus the same arrangement is used to examine both beams.

Once established, OIL is stable and reliable. Slave reproduces the spectral properties of the seed and amplifies its power to < 50 mW. The locking range of ~ 2 GHz spans the MHF range of the ECDL. In earlier investigations, while using ECDL (fitted with DL of other make) with ~ 5 GHz MHF tunability and applying additional synchronization between the master PZT voltage and the slave bias current, we were able to observe stable OIL over this broader range.

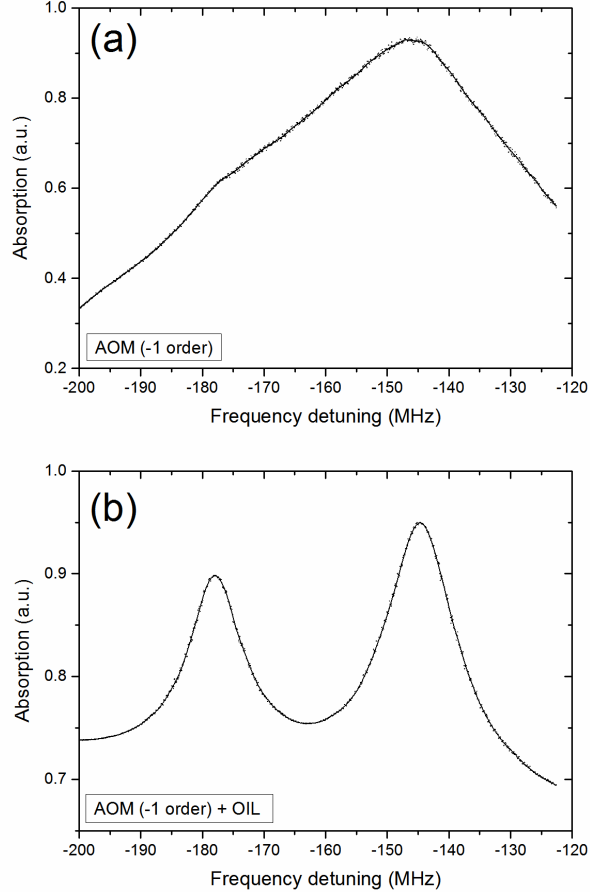

**Fig. 4** Saturated-absorption sub-Doppler features of the $^{85}$Rb D2 transition. Spectrum taken by scanning the laser frequency with the double-pass AOM, aligned in the -1 order. Detuning is measured relative to the DAVLL locked frequency of the master laser. The crossover resonances F = 3 → F'= (2,4) (left peak) and F = 3 → F'= (3,4) (right peak) separated by 31.5 MHz are shown. Note that the lines are power-broadened and the wide Doppler pedestal is not subtracted. The laser beam is taken: in (a) immediately after AOM. Distortion of the spectrum is caused here by a strong dependence of the AOM diffraction efficiency on the carrier frequency; in (b) after the OIL stage. Spectrum regains its regular shape. The optical limiting of the OIL amplifier is manifested.

Figs. 2(b) and 3(b) depict the normalized response of the OIL stage injected with the AOM detuned beam (double-pass AOM aligned in -1 and +1 order, respectively) as a function of the RF carrier frequency. Note that different vertical scales are applied for curves (a) and (b) in each Figure. Two factors influence observations: (i) input power of variable dynamic range ~ 1:2 and (ii) optical frequency detuning variable in the range of ~ 70 MHz.

A nearly flat response is observed. The injection locked output power is hardly affected by significant changes in the seed intensity, the advantageous fact known also from earlier studies, e.g., [90,94,95]. The OIL stage, thus, acts as an optical limiting amplifier, effectively suppressing variations in optical input signal strength. On the other hand, changing detuning leads to a frequency-dependent, weak but noticeable (~ 0.5% at most), effect. As can be seen, the power changes nearly monotonically, increasing (decreasing) in Fig. 2(b) (in Fig. 3(b), respectively) with the RF carrier frequency. This behavior is characteristic for a semiconductor OIL amplifier because its gain is detuning dependent, with the output power increase as the signal tunes red [69,88,90].

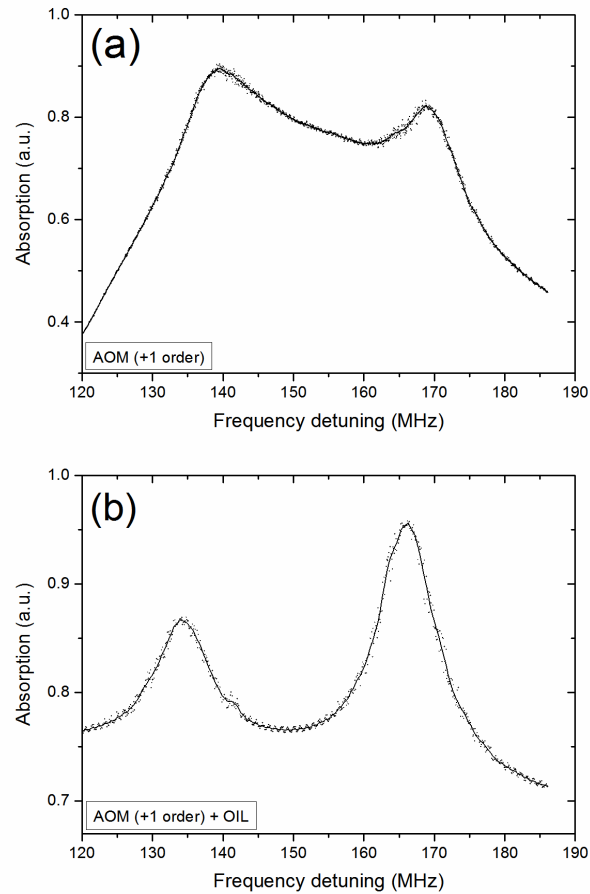

**Fig. 5** The same as in Fig. 4 but here the double-pass AOM is aligned in the +1 order. The applied laser beam is taken: (a) immediately after AOM; (b) after the OIL stage. The optical limiting of the OIL amplifier is manifested.

(Notes: (i) It is not clear for us why the monotonicity breaks down close to the low (high) RF limit in Fig. 2(b) (in Fig. 3(b), respectively), which corresponds to the blue limit of the optical frequency band. (ii) Stability of the injection-locked slave after injection with the AOM detuned seed was reported in [96].)

OIL retains the spectral and spatial properties of the seed beam. Acting as a highly saturated amplifier, it provides nearly constant output power throughout the AOM tuning range. Strong variation of the signal (Figs. 2(a) and 3(a)) is entirely suppressed in the injection-locked output, its residual dependence on detuning is very weak and (nearly) monotonic in the applied limited bandwidth of 70 MHz (Figs. 2(b) and 3(b)). OIL acts as an optical limiting amplifier that stabilizes the output power at a predefined value for different input signal strengths.

Another advantage of the OIL stage is the compensation of pointing fluctuations of the seed beam, which arise during frequency tuning with AOM or PZT. It allows a stable coupling of the OIL amplified beam into the small aperture of the next stage amplifier or of the optical fiber.

To illustrate the benefits of OIL for spectra registration, the sub-Doppler SAS features of the $^{85}$Rb D2 transition are shown in Figs. 4 and 5. The crossover resonances F=3 → F'=(2,4) (left peak) and F=3 → F'= (3,4) (right peak) separated by 31.5 MHz are given. Spectra are taken under various conditions using the system in Fig. 1. Frequency detuning is accomplished with a double-pass AOM and measured relative to the null point defined by the DAVLL locked frequency of the master laser. AOM is aligned in -1 order (in Fig. 4) or in +1 order (in Fig. 5), hence the horizontal scales and their null points are different in both cases. The spectra in Figs. 4(a) and 5(a) are taken with the beam directly after the AOM. They are heavily distorted (line shapes are deformed and their centers shifted). In contrary, spectra in Figs. 4(b) and 5(b) are taken with the beam after the OIL amplifier. Due to the deep saturation, the OIL stage compensates for signal variations (optical limitation), therefore undistorted spectra can be observed. Note that in the spectra the background of the Doppler broadened pedestal is not subtracted, and the linewidths are power-broadened. The weak dependence of the intensity on detuning (as seen in Figs. 2(b) and 3(b)) has no noticeable effect. The subtle modulation structure visible in Fig. 5(b) results from mechanical disturbances during the measurement time (~ 0.5 s). The spectra demonstrate the suitability of our laser system for precise measurements.

## 2.5 TA power stage

Higher powers are achieved by injecting the OIL amplified beam into the TA (MOPA configuration). Contrary to the OIL where the input beam frequency and working parameters of the amplifier are strongly interrelated the TA is not critically sensitive to the bias current and temperature and amplifies frequencies across its broad gain profile. We use the m2k Laser TA-0780-1000-CM device of a 780 nm center frequency, 1W specified output power, 2 mm length, 20–30 mW suggested seed power, 2.1 A operation current, 767–787 nm gain bandwidth, TM polarization (i.e., parallel to the fast axis), and $M^2 = 1.3$ beam quality factor. A bare, and thus vulnerable, semiconductor structure fixed onto the C-mount consists of two sections: a short and narrow pre-amplifying ridge-waveguide acting also as a mode filter, and a long power amplifying tapered section. The size of exit aperture is 205 μm by 1.3 μm. The size of the entrance aperture is not given (~ 4 μm width is expected) but its aspect ratio is small. Both apertures are quality (< 0.01%) AR-coated. We ordered two copies of this amplifier and were kindly offered one extra for testing purposes. Worse spatial properties of the beam are expected for this test diode. Originally we planned to use it only during the system assemblage and first tests when the risk of failure seemed to be particularly high. However, since the diode behaved better than expected we extended its utilization. Results presented below concern this particular TA chip. TA handling must be careful to protect the chip against static discharge, mechanical damage and contamination. Various mechanical arrangements of the TA heads have been reported [68,73-77,97], in some cases detailed technical drawings were provided. Our arrangement is inspired by the design of [98] with many changes applied, e.g., in the beam collimation system. The TA's C-mount (anode) is bolted, securing good electrical and thermal contact, into the slot in the precision-machined copper block and heat sink. The block is fixed firmly with screws inside the rectangular (40 x 60 x 60 $mm^3$) aluminum made enclosure. The latter is supplied with two openings of 20 mm diameter centered along the optical axis for input and output beams. Inside the enclosure, a small PCB is mounted containing a Schottky diode (BAT 85) to protect TA against reverse voltage above 0.3 V and a miniature relay (Axicom IM03) (with another Schottky diode as a shunt) to short the device when the current is off. It also includes wiring for TA power and control, and a miniature 6-pin flat connector that goes through the front of the housing. To reduce strain on the TA structure, its cathode connection, a delicate wire flag attached to the ribbon, is cautiously soldered with a short lead to the protective board. Another wire connects the anode (C-mount) and the copper block to the board. To improve heat transfer and avoid the use of heat-conducting paste, thin indium foil is applied between the C-mount and the copper block and on other relevant surfaces. The temperature (20 ºC recommended) is stabilized with the LM35 sensor glued into a small hole in the copper block close to the TA chip, TEC (40 x 40 $mm^2$) squeezed between the rear wall of the housing and a massive aluminum made heat sink, and the temperature controller of the type described earlier. The heat sink is fixed onto the optical board for stability and heat dissipation. Because of moderate power no water cooling is applied. Current is supplied via a low-noise 2.5 A controller, designed and built in our lab, with functionalities similar to those used in the previously described high precision power supply.

To simplify the system, commercially available components have been used. The vertically polarized input beam is focused on the TA entrance facet by an aspheric AR-coated lens, the same type as that used in the slave. The exit beam is shaped with a two-lens set consisting of an aspheric lens (Thorlabs C330TME-B, f=3.1 mm, NA=0.68, AR-coated) and a cylindrical lens to account for TA astigmatism. Both aspheric lenses are held in threaded adapters (Thorlabs S05TM09) and in adjustable lens tubes (Thorlabs SM05V05), which are mounted in biaxial positioning lens holders (Thorlabs SCP05) providing transverse adjustment of the collimators. The lens holders, in turn, are fixed in their mirror mounts (Thorlabs KS1), which are then attached to both ends of a solid aluminum plate (150 x 40 x 15 $mm^3$). The plate is rigidly attached to the bottom of the TA housing. Conically shaped nitrile rubber gaskets are fitted between lens tubes and the corresponding openings in the aluminum enclosure, providing tight and reliable yet elastic seals to protect TA against dust ingress. The fully assembled TA head is shown in Fig. 6. Its ultimate performance depends on careful alignment. Fine threaded adjusters and a spanner wrench allow precise positioning of collimators in all axes. The TA power supply was not fitted with the suitable functionality, thus the approach described earlier for the OIL laser was not applicable. Guidelines on the TA alignment procedure can be found, e.g., in [77].

TA beam is characterized by widely differing divergences in directions parallel (slow axis) and perpendicular (fast axis) to the waveguide plane (full angles for 95% enclosed power amount 17° and 45°, respectively) and strong astigmatism in the slow axis. While the waist of the fast axis is at the exit facet, the (virtual) waist of the slow axis is located in the plane of the waveguide approximately at a distance $L/n_g$ behind the facet (assuming $L \approx 1.5$ mm, $n_g \approx 3$, the distance is ~ 0.5 mm; also depends on the current), where $L$ is the length of the tapered section and $n_g$ the effective refractive index [61]. Therefore, in addition to an aspheric lens placed by one focal length from the exit facet and providing collimation in the fast axis, a cylindrical AR-coated lens was used in the slow axis to correct this distortion. Its focal length f = 75 mm was chosen on account of mechanical constraints of the arrangement. The resulting ~3 mm wide beam is elongated horizontally with the calculated aspect ratio ~5:3. A double-stage isolator (Gsänger DLI-1, > 60dB, aperture 5 mm, transmission ~ 85%) is placed behind the cylindrical lens to prevent back reflection to TA that could easily damage its structure. If necessary, further beam shaping can be performed with a pair of anamorphic prisms (to improve circularization) or a telescope (to change diameter). The spatial quality of the output beam is inferior to that of the injection beam. While in the fast axis the intensity distribution nearly follows the Gaussian profile, in the slow axis it has a more irregular pattern. For the spatial mode cleaning a pinhole or coupling into fiber can be utilized.

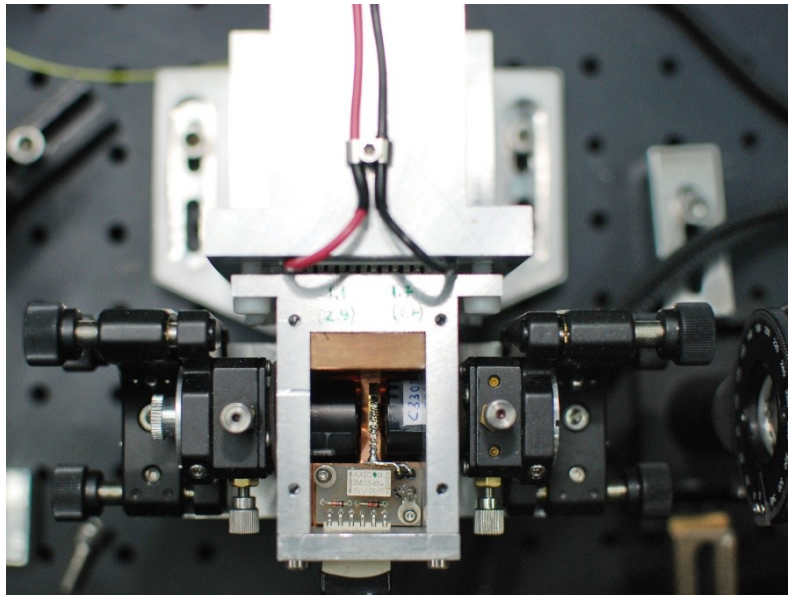

**Fig. 6** A fully assembled TA head with the lid open. TA chip (not visible) fixed with a screw to the copper block (and a heat sink) is situated between the input and output collimator tubes. The latter are mounted in their respective xy translation stages and kinematic mirror mounts to allow precise alignment of the beams. A small PCB contains protection diodes and a relay, power wiring, and a miniature connector. The TA head is attached with plastic screws to a large aluminum base and a cooling element is installed between them. The laser beam propagates from left to right.

The TA output power characteristics is given in Fig. 7 in two different projections, the input (output) power is measured before the collimator (after the isolator, respectively). The seed power is varied by means of a half-wave plate mounted in the motorized rotation stage (Thorlabs PRM1Z8) and the PCB4 placed just behind it. Fig. 7(a) shows the dependence of the output power on the seed power (in the range from 0.4 mW to 40 mW) for the amplifier currents in the range from 0.8 A to 2.2 A. (Note: currents above 1.5 A are not recommended in unseeded TA.) For each current value, the power increases with the increase in seed power, but the growth rate gradually decreases, which proves a partially saturated gain. For smaller currents, saturation is more pronounced and starts at lower input powers. Fig. 7(b) shows the dependence of the output power on the current for selected seed powers. There is a pronounced increase in power when the current exceeds the threshold value of ~ 0.6 – 0.8 A. For example, at a seed level of 12 mW (20 mW) and currents in the range of 1.6 – 2.2 A, an almost linear growth is observed with the slope coefficient of 0.45 W/A (0.46 W/A). As can be seen, as the input power increases, the slope increases and the threshold current decreases. Results of [77] for TA of the same type in alike conditions (12.6 mW seed, output power measured after the isolator) provide 0.41 W/A. In typical operation conditions of 1.9 A bias current and 12 mW (20 mW) seed power we observe an output of 404 mW (467 mW), which translates into amplification factor of 33.7 (23.3) or gain of 15.3 dB (13.7 dB, respectively). In [77] with 12.6 mW seed and 1.9 A current a similar 430 mW output power and 15.3 dB gain (or 565 mW and 16.5 dB gain, when measured before the isolator) is reported, which compares favorably with the manufacturer data. Thus, our power data seem to agree well with those in [77] for the same TA. However, as seen in Fig. 7 TA does not saturate fully despite high (up to 40 mW) seed power, also the output power is slightly below expected levels. The above may suggest that the TA characteristics have not reached optimum. The obvious reason of lower power are losses in optical components (15% in the isolator alone). Unoptimized coupling efficiency can contribute to both reduced saturation and reduced output power. Several factors can be distinguished here, such as: inaccurate focusing of the seed, inadequate matching of its waist with the TA input facet or offset between the beam axis and the TA axis, which all affect the power injected into the narrow waveguide [97]. The efficiency of injection also depends on the seed beam polarization direction, which must coincide (within a few degrees) with the direction specified for TA. However, its control is ensured with the half-wave plate placed in front of the collimator (and tilted to prevent feedback). We conclude that during the described tests, the waveguide power coupling was not fully optimized, which reduced the efficiency of TA. Some performance loss can also be attributed to the characteristics of our particular TA (test diode).

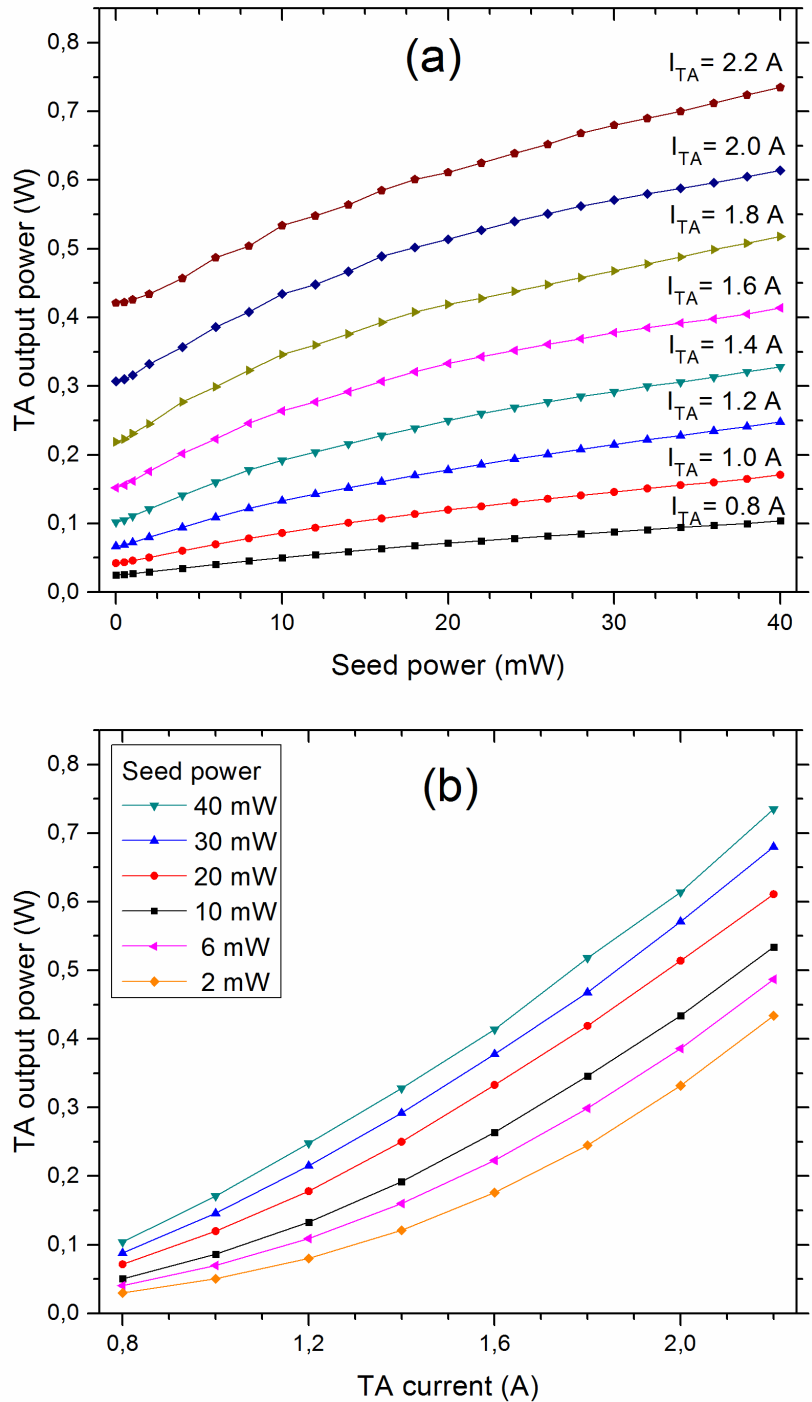

**Fig. 7** Power characteristics of the TA stage. (a) Output power as a function of input (seed) power for different operating currents. The lowest (highest) input power is 0.4 mW (40 mW). (b) Output power as a function of operating current for selected values of the seed power.

TA maintains spectral quality of the input beam [73]. Our analysis with an FP spectrum analyzer and a SAS spectrometer showed no line broadening. However, in addition to the narrow coherent peak at the output, there is a broadband background due to amplified spontaneous emission (ASE). The ASE power and its fraction in the total output varies with the TA operating conditions. It is highest at low seed levels and high injection currents. For the unseeded TA we observed ASE power of 5.6 mW (35.8 mW) at 0.5 A (1.5 A, respectively). No attempt was made to assess the contribution of ASE to the results of Fig. 7. Under operating conditions, the gain is depleted by the amplified signal and the ASE is attenuated. As shown in [68,76], the ASE fraction decreases sharply with increasing seed power, as the gain saturates. At full saturation, fractions of up to a few percent (exact amount being temperature dependent) [68] or ~ 10% [76] are reported for various 780 nm devices. Since the power spectral density at the narrow coherent peak is several orders of magnitude stronger than that of the ASE pedestal, in some important applications (e.g., in laser cooling) the ASE impact is irrelevant [68,77]. Filtering with optical fiber is an effective approach to ASE suppression and improving the spatial quality of the beam. The single mode (SM) fiber coupling efficiencies of 25-35% [77], 46% [68], 45-55% [75] or >50% [97] are reported for TA beams. For single mode polarization maintaining (SM/PM) fibers lower efficiencies are expected.

The steering mirrors (M10, M11) and the half-wave plate allow fiber coupling. The half-wave plate is used to match the beam polarization to the optical axis of the SM/PM fiber. To prevent back reflections a fiber with tilted facets is used. Then, the spatially and spectrally purified Gaussian beam after the fiber can be applied in the experiment. Since the TA beam was not fully optimized, the SM/PM fiber coupling was inefficient and therefore we skip our preliminary results.

In its present form, the TA stage may require some small adjustments each time the beam is to be used, before thermal equilibrium is reached or occasionally to compensate for long term drifts. A more careful alignment on the input side would maximize the seed beam mode matching and increase the TA power efficiency. A better circularization and narrower waist of the output beam is required for an efficient coupling to the SM/PM fiber. Thus a more compact collimation system and a better beam shaping is considered. The entire amplifying system is mounted on a 60 x 120 cm$^2$ optical breadboard, which rests over a layer of elastic foam to isolate vibrations.

## 3   Conclusions and outlook

A simple three-stage amplifying tunable diode laser system operating at 780 nm capable of driving the D2 transitions of Rb was developed. The essential components, in particular the lasers and the laser controllers, are designed and built in our lab. The sequence comprises: frequency stabilized ECDL master laser, double-pass AOM frequency shifter, OIL amplifier and TA power unit. The basic parameters of the system are as follows. The time-averaged linewidth of the master laser and the entire system is ~ 0.6 MHz and its MHF tunability is ~ 2 GHz. We applied DAVLL [51,55,56], a simple, modulation-free long-term frequency stabilization scheme, based on an induced circular dichroism, that requires a few rather inexpensive components. It offers a wide capture range and a broad continuous tuning range of 400 MHz, within the Rb D2 transition, with a rather small drift < 2 MHz/h. The applied AOM model ensures precise frequency tunability in the 70 MHz bandwidth. Although narrow, this bandwidth covers the interesting spectral features of the Rb spectrum. By using a different AOM, one can easily expand the tuning range. The beam is amplified in the OIL stage and then in the TA power stage to over 700 mW, while preserving the master linewidth.

Properties of the OIL stage have been analyzed. OIL acts as a saturated optical limiting amplifier that stabilizes the output power at a predefined value for different input signal strengths. Optical limitation systems based on various mechanisms are of interest in many applications [90,94,95,99]. We are not aware of any report of this property, directly applicable to the registration of spectra in an OIL based system using AOM frequency sweep. Therefore, we found it useful to demonstrate it with the aid of example spectra. The diffraction efficiency of AOM is strongly frequency dependent, which inevitably degrades the spectrum taken directly, unless the OIL effectively (in our case with an accuracy better than 0.5%) inhibits the intensity variations. Another advantage of the OIL stage is that it compensates for slight pointing fluctuations of the seed beam, while its frequency is swept by the AOM (or PZT). There is no sign of beam shift or deflection after the OIL stage. Thus, it allows the OIL amplified beam to be stably coupled to the small aperture of the next stage amplifier or optical fiber.

We have also developed an approach to optimizing injection efficiency of the seed beam into the slave cavity, where the light-induced increase in voltage at the junction, operated in a photovoltaic (zero-bias) mode, is directly measured with a dedicated add-on function built into the laser controller, without the need to disconnect or re-wire the DL.

As compared to BAL based systems the optical arrangement of TA systems is simpler, beam quality is better and usually less input power is required [61]. (It should be noted, however, that as an alternative to the amplifying laser systems, a high performance 1000 mW single-stage tunable ECDL based on a tapered laser was presented in, e.g., [24].) Our arrangement employs popular DLs without AR-coating, while TA chips are common and available for many wavelengths, in addition, the entire system requires relatively few optical components. A simple, reliable and easy-to-operate 780 nm diode laser system of significant power has been developed, we expect the details presented and discussed to be useful to other users.